\newif\ifjournal
\journalfalse
\ifjournal
  \documentclass{mrlart4}

  \address{Department of Mathematical Sciences, University of Alaska Fairbanks, Fairbanks, Alaska 99775}
  \email{david.maxwell@uaf.edu}
\else
\documentclass{article}
\usepackage{times}
\usepackage{geometry}
\date{April 5, 2008}
\fi
\usepackage{amsmath,color,amsthm}
\ifjournal
\title[Vacuum Einstein constraints]{A class of solutions of the vacuum Einstein constraint equations with freely specified mean curvature}
\else
\title{A class of solutions of the vacuum Einstein constraint\\ equations with freely specified mean curvature}
\fi
\author{David Maxwell}

\DeclareMathOperator{\ck}{\bf L}
\DeclareMathOperator{\Lap}{\Delta}
\DeclareMathOperator{\Lichop}{\mathcal L}
\DeclareMathOperator{\Mop}{\mathcal W}

\DeclareMathOperator{\compop}{\mathcal N}

\DeclareMathOperator{\Vol}{\rm Vol}
\DeclareMathOperator{\tr}{\rm tr}
\newcommand{\dV}{\; dV}
\newcommand{\abs}[1]{\left|#1\right|}
\newcommand{\norm}[1]{\left|\left|#1\right|\right|}
\renewcommand{\div}{\mathop{\rm div}\nolimits}
\newcommand{\yamabe}{{\mathcal Y}}

\newcommand{\ra}{\rightarrow}
\newcommand{\dom}{{\mathcal D}}
\let\Lop\ck
\def\ip<#1,#2>{\left<#1,#2\right>}
\begin{document}
\newtheorem{theorem}{Theorem}
\newtheorem{proposition}{Proposition}
\newtheorem{corollary}{Corollary}
\newtheorem{lemma}{Lemma}
\maketitle
\begin{abstract}
We give a sufficient condition, with no restrictions on the mean curvature,
under which the conformal method can be used to generate solutions of the 
vacuum Einstein constraint equations on compact manifolds.
The condition requires a so-called global supersolution but
does not require a global subsolution.  
As a consequence, we construct a class 
of solutions of the vacuum Einstein constraint equations
with freely specified mean curvature, extending 
a recent result \cite{Holstetal07} which constructed 
similar solutions in the presence of matter. We give a second
proof of this result showing that vacuum solutions can be obtained as a limit of
\cite{Holstetal07} non-vacuum solutions.
Our principal existence theorem is of independent interest in the near-CMC case,
where it simplifies previously known hypotheses required for existence.
\end{abstract}

\section{Introduction}
The Cauchy problem of general relativity requires initial data (a metric and
a second fundamental form defined on a 3-manifold) that satisfy 
a system of nonlinear PDEs known as the Einstein constraint equations.  The constraint
equations admit many solutions (permitting the specification of different initial conditions)
and it is important to understand the structure of the
set of all possible initial data on a given manifold.  Various approaches have been given for 
constructing solutions including parabolic methods \cite{BartnikQS} and gluing
constructions  \cite{corvino} \cite{InitialDataEngineering}.  From the point of view
of classifying the set of all possible solutions, the most fruitful technique has
been the conformal method initiated by Lichnerowicz \cite{lich-conformal} and extended 
by Choquet-Bruhat and York \cite{cb-york-held}.

In the conformal method, one specifies the conformal class of the initial metric,
a piece of the second fundamental form corresponding to part of the time derivative 
of the conformal class, and the trace of the second fundamental form (i.e. the
mean curvature). One then seeks a solution of the constraint equations matching this data.  
For constant mean curvature (CMC) data, this approach has lead to a complete 
classification of solutions on compact \cite{cb-york-held} \cite{const-compact}, 
asymptotically Euclidean \cite{cantor-cond} \cite{lap-asympt-flat} (with a correction
in \cite{maxwell-ahc}), and asymptotically hyperbolic 
\cite{Anderssonetal92} \cite{Anderssonetal96} manifolds.
On the other hand, we have very few results concerning non-CMC solutions,
and most of these are perturbative. Near-CMC solutions have been constructed
on compact \cite{BruhatEtAl92} \cite{const-non-cmc-compact} \cite{Isenbergetal07} \cite{Holstetal07}
and asymptotically Euclidean \cite{BruhatAH} \cite{const-asympt-flat} manifolds, and we have
a near-CMC non-existence theorem for certain data \cite{IsenbergOMurch04}.  It is remarkable, however, that despite
the success of the conformal method in the CMC case, very little is known
about the construction of solutions in the absence of restrictions on the mean curvature.

An important recent result of Holst, Nagy, and Tsogtgerel \cite{Holstetal07} (see also the summary in
\cite{Holstetal08}) gives the 
first construction, using the conformal method, of a class of initial data without a near-CMC hypothesis.  
The authors of that paper show that solutions of the constraint equations on compact manifolds
can be constructed using the conformal method when global sub- and supersolutions (defined 
in Section \ref{sec:conformal})
can be found.  For Yamabe-positive metrics, for non-vanishing matter fields, and
under a certain smallness condition not involving the mean curvature,
\cite{Holstetal07} provides a global subsolution/supersolution pair
that does not have a near-CMC hypothesis and hence yields the existence of 
certain far-from-CMC solutions.

It is natural to ask if this far-from-CMC construction can
be extended to the vacuum case.  The global supersolution 
of \cite{Holstetal07} (hereafter the HNT global supersolution) is also applicable in vacuum,
and indeed requires that the matter fields, if present, be weak.  The 
corresponding HNT global subsolution, however,  requires the presence of matter.  
It is not unusual for the conformal method to require non-vanishing conditions on parts
of the specified data, so it was conceivable that the non-vacuum hypothesis
was necessary.

In this paper we show that this is not the case, and that the conformal method can be
used to construct a corresponding set of vacuum solutions.  We give two proofs 
of this fact.  First, we prove that solutions exist, under certain mild technical conditions,
whenever a global supersolution can be found (Theorem \ref{thm:main}).
The proof relies on an a-priori estimate (Proposition \ref{prop:yposlow}) that replaces the
need for a global subsolution.  Hence the HNT global supersolution alone is sufficient to 
deduce the existence of solutions via the conformal method, and we obtain vacuum far-from-CMC
solutions.  The second proof considers a sequence of HNT non-vacuum solutions where the matter fields
are converging to zero.  Again, a lower bound (Proposition \ref{prop:seqlow}) is found for the sequence
and is used to obtain a corresponding subsequence converging to a vacuum far-from-CMC solutions.  
The key steps in both proofs rely on a technique from \cite{Maxwell05} for constructing subsolutions.

\subsection{The conformal method}\label{sec:conformal}
On a given smooth $3$-manifold $M$, the Einstein constraint equations  
for a metric $\bar g$ and a symmetric $(0,2)$-tensor $\bar K$ are
\begin{equation}\label{constraints}
\begin{aligned}
R_{\bar g}-\abs{\bar K}_{\bar g}^2+ \tr_{\bar g} \bar K ^2 &= 2\bar \rho\\
\div_{\bar g} \bar K -d\,\tr_{\bar g} \bar K &= \bar J,
\end{aligned}
\end{equation}
where $R_{\bar g}$ is the scalar curvature of $\bar g$, $\bar \rho$ is the matter density, and $\bar J$
is the momentum density.  We are primarily interested in the vacuum case where $\bar\rho \equiv 0$ and
$\bar J\equiv 0$.
 
Data for the vacuum conformal method on a compact smooth manifold $M$
consists of a Riemannian metric $g$ specifying a conformal class, a 
transverse traceless (i.e symmetric, trace-free and divergence-free) $(0,2)$-tensor $\sigma$ specifying
part of the time derivative of the conformal class, and a scalar function
$\tau$ specifying the mean curvature.
We seek a solution $(\bar g, \bar K)$ of the constraint equations 
of the form
\begin{equation}\label{reconst}
\begin{aligned}
\bar g &= \phi^{4} g\\
\bar K &= \phi^{-2}\left(\sigma+\ck W\right)+{\tau\over 3}\tilde g.
\end{aligned}
\end{equation}
In equations \eqref{reconst} the unknowns are a positive function $\phi$ and a vector field $W$,
while $\ck$ is the conformal Killing operator defined by
\begin{equation}
{\ck W}_{ab} = \nabla_a W_b + \nabla_bW_a - {2\over 3} \div W g_{ab}.
\end{equation}
If matter is present, it can be specified by scaled sources $\rho$ and $J$ which 
are conformally related to $\bar \rho$ and $\bar J$ by $\bar \rho=\phi^{-8}\rho$ 
and $\bar J = \phi^{-6}J$.

It follows that $\bar g$ and $\bar K$ solve the vacuum constraint equations
so long as
\begin{align}\label{sys-lich}
-8\Lap \phi + R_g \phi &= -{2\over 3}\tau^2\phi^{5} + \abs{\sigma+\ck W}^2\phi^{-7}\\
\div \ck W &={2\over 3} \phi^{-6} \;d\tau. \label{sys-momentum}
\end{align}
If matter is present we must add the terms $2\rho\phi^{-3}$ and $J$ to the right-hand 
sides of \eqref{sys-lich} and \eqref{sys-momentum} respectively.  The operator
$\div \ck$ is the vector Laplacian, and hence these equations are a coupled nonlinear
elliptic system for $\phi$ and $W$.

If $\tau$ is constant then equation \eqref{sys-momentum} has a trivial solution 
and the problem reduces to an analysis of the Lichnerowicz equation
\eqref{sys-lich}.  One 
technique for finding solutions of the Lichnerowicz equation is via 
the method of sub- and supersolutions, which was used previously 
in the work of Kazden and Warner \cite{KazdanWarner75} on the
prescribed scalar curvature problem.  Isenberg \cite{const-compact}
used this method
to complete the classification of CMC solutions on compact manifolds.
A generalization of the method applies
in the non-CMC setting as well, and we review the terminology now.

Consider the equation
\begin{equation}\label{lichfirst}
-8\Lap \phi + R \phi = -{2\over 3}\tau^2\phi^{5} + \abs{\beta}^2\phi^{-7} 
\end{equation}
where $\beta$ is a symmetric $(0,2)$-tensor. We say $\phi_+$ is a {\bf supersolution}
of \eqref{lichfirst} if
$$
-8\Lap \phi_+ + R \phi_+ \ge  - {2\over 3}\tau^2\phi_+^{5}+\abs{\beta}^2\phi_+^{-7}.
$$
A {\bf subsolution} is defined similarly with the inequality reversed.  

For the coupled system, we follow \cite{Holstetal07} and define global subsolutions
and global supersolutions as follows.  Given a function $\phi$, let $W_\phi$ be
the corresponding solution of \eqref{sys-momentum}.  We say $\phi_+$ is a {\bf global supersolution}
if whenever $0<\phi\le \phi_+$, then
$$
-8\Lap \phi_+ + R \phi_+ \ge - {2\over 3}\tau^2\phi_+^{5}+\abs{\sigma+\ck W_\phi}^2\phi_+^{-7} .
$$
We say $\phi_->0$ is a {\bf global subsolution} if whenever $\phi\ge \phi_-$, then
$$
-8\Lap \phi_- + R \phi_- \le  - {2\over 3}\tau^2\phi_-^{5} +\abs{\sigma+\ck W_\phi}^2\phi_+^{-7}.
$$
The existence result of \cite{Holstetal07} states that if $\phi_-\le \phi_+$ are global sub- and
supersolutions, then there exists a solution $(\phi,W)$ of system 
\eqref{sys-lich}--\eqref{sys-momentum} such that $\phi_-\le\phi\le\phi_+$.  The authors
of that paper also present a number of global sub- and supersolution pairs, including one 
that is used to construct far-from-CMC solutions.

\subsection{Summary of results}

Our primary result concerning the solution of system \eqref{sys-lich}--\eqref{sys-momentum}
has three cases depending on the Yamabe invariant $\yamabe_g$ of the metric.
Recall that 
$$
\yamabe_g =  \inf_{f\in C^{\infty}(M)\atop f\not\equiv0} 
{\int_M 8\abs{\nabla f}^2_g + R_g f^2\dV_g\over ||f||_{L^{6}}^2}.
$$
(Our notation for $L^p$ spaces and Sobolev spaces $W^{k,p}$ follows that of
\cite{yamabe-problem} with the additional convention that subspaces of positive functions are
indicated by a subscript $+$.)

\begin{theorem}\label{thm:main} Let $g\in W^{2,p}$ with $p>3$ be a metric on a smooth, compact 3-manifold.  Suppose $g$ has no conformal Killing fields and that
one of the following conditions holds for 
a transverse traceless tensor $\sigma\in W^{1,p}$ and a function $\tau\in W^{1,p}$.
	\begin{enumerate}
		\item $\yamabe_g>0$, $\sigma\not\equiv 0$,
		\item $\yamabe_g=0$, $\sigma\not\equiv 0$, $\tau\not\equiv 0$
		\item $\yamabe_{g} < 0$ and there exists $\hat g$ in the conformal class of $g$ such that
	  $R_{\hat g} = -{2\over 3}\tau^2$.
	\end{enumerate}
	If $\phi_+\in W^{2,p}_+$ is a global supersolution for $(g,\sigma,\tau)$, then 
	there exists a solution $(\phi,W)\in W^{2,p}_+\times W^{2,p}$ of 
	system \eqref{sys-lich}--\eqref{sys-momentum} such that $\phi\le \phi_+$.
\end{theorem}

The new results of Theorem \ref{thm:main} are Cases 1 and 2; Case 3 
can be deduced from the existence of a global subsolution found in \cite{Holstetal07}.
Note that Theorem \ref{thm:main} is only an existence theorem. It is not known if the
solutions provided by Theorem \ref{thm:main} are unique.

If $\tau$ is constant then the conditions of the three cases 
reduce to precisely the same conditions under which CMC solutions of the constraints can be found (aside from
one additional singular case $\yamabe_g=0$, $\sigma\equiv0$, $\tau\equiv0$).  
The hypothesis on $\tau$ in Case 3 is necessary for Yamabe-negative metrics since it is 
needed for solutions of the Lichneowicz equation to exist \cite{Maxwell05}.  It is not known if the condition $\sigma\not\equiv 0$ in Cases 1--2 is necessary.  However, it was
proved in \cite{IsenbergOMurch04} that if $\yamabe_g\ge 0$ and if $\sigma\equiv 0$, then
there do not exist near-CMC solutions of \eqref{sys-lich}--\eqref{sys-momentum} 
unless $\yamabe_g=0$ and $\tau\equiv 0$ also. Hence some condition involving $\sigma$ 
(and possibly also $\tau$) must be required.  
The hypothesis in Case 2 that $\tau\not\equiv 0$ (if $\sigma\not\equiv 0$) can
be shown to be necessary -- otherwise the metric would be be Yamabe-positive.

Our first application of Theorem \ref{thm:main} is to the HNT supersolution,
which exists under the following hypotheses.
\begin{proposition}[{\cite{Holstetal07}}] \label{prop:HNTsuper}
Suppose $g\in W^{2,p}$ with $p>3$, and that
$\yamabe_g>0$, $\tau\in W^{1,p}$, and $\sigma \in W^{1,p}$.
If\; $||\sigma||_{\infty}$ is sufficiently small, then
there exists a global supersolution of \eqref{sys-lich}--\eqref{sys-momentum}.
\end{proposition}
A proof of Proposition \ref{prop:HNTsuper} can be found in Section \ref{sec:mapping}.
From Theorem \ref{thm:main} and Proposition \ref{prop:HNTsuper} 
we immediately obtain the following result, which is the primary aim
of this paper.
\begin{corollary}\label{cor:nonvac}
Let $g\in W^{2,p}$ with $p>3$ be a Yamabe-positive metric on a smooth compact 3-manifold.
Suppose $g$ has no conformal Killing fields, $\sigma\in W^{1,p}$ is a transverse traceless
tensor, and $\tau\in W^{1,p}$. If $\sigma\not\equiv 0$ and if\; $||\sigma||_\infty$ is
sufficiently small,
then there exists a solution $(\phi,W)\in W^{2,p}_+\times W^{2,p}$  of system \eqref{sys-lich}--\eqref{sys-momentum}.
\end{corollary}
We also provide a second proof of Corollary \ref{cor:nonvac} in Section \ref{sec:sequence} 
that is independent of Theorem \ref{thm:main} but instead uses a sequence of HNT
non-vacuum solutions.

Theorem \ref{thm:main} permits a strengthening of the current existence theory for
near-CMC data inasmuch as it removes conditions in current theorems required to find
subsolutions.  In \cite{Isenbergetal07} the authors present an existence and uniqueness
theorem for Yamabe non-negative metrics. They present constant global supersolutions 
so long as
\begin{equation}\label{smallmc}
\frac{\max(\abs{ d\tau})}{\min(\abs{\tau})}\;\;\text{is sufficiently small},
\end{equation}
and also assume non-scale invariant conditions on the size of 
$\abs{d \tau}$ to obtain subsolutions and to obtain uniqueness. Previously \cite{const-non-cmc-compact}
gave a similar proof for Yamabe-negative metrics, again presenting a global
supersolution under condition \eqref{smallmc} and making an additional assumption
about the absolute size of $\abs{d\tau}$.
\cite{Holstetal07} provides a global supersolution under a near-CMC condition similar
to \eqref{smallmc}, but also requires in the Yamabe-nonnegative case either 
a non-vacuum hypothesis or that $\min{\abs{\sigma}}$ is sufficiently large 
to obtain a subsolution.  
Using Theorem \ref{thm:main} we have the following simplified existence result.
\begin{corollary}\label{cor:nearcmc}
Let the conditions of one of the cases of Theorem \ref{thm:main} hold.	If
$$
\frac{\max(\abs{ d\tau})}{\min(\abs{\tau})}
$$
is sufficiently small, 
then there exists a solution $(\phi,W)$ of system \eqref{sys-lich}--\eqref{sys-momentum}.
\end{corollary}

The utility of Theorem \ref{thm:main} is limited to 
cases where a global supersolution can be found. 
It is not known if the converse
of Theorem \ref{thm:main} is true; in particular, given a solution of system \eqref{sys-lich}--\eqref{sys-momentum} 
it is not known if there also exists a corresponding
global supersolution.\footnote{Given a solution $(\phi, W)$ of system \eqref{sys-lich}--\eqref{sys-momentum},
it is not clear if $\phi$ itself is a global supersolution.  Although $\phi$ is, for 
one particular $W$, a solution and hence a supersolution of \eqref{sys-lich},
it is not apparent that it is a supersolution for the whole class of vector fields $W$ 
required for it to be a global supersolution.}
Nevertheless, Theorem \ref{thm:main} makes clear that any future advances 
in the existence theory of non-CMC initial data
using the method of sub- and supersolutions need only focus on supersolutions.

In the following, Sections \ref{sec:lich} and \ref{sec:momentum} provide a summary
of the basic results we require in the analysis of equations \eqref{sys-lich} and
\eqref{sys-momentum} respectively. Section \ref{sec:coupled} is devoted to 
the proof of Theorem \ref{thm:main}, which is obtained using the Schauder fixed point
theorem in an approach similar to one outlined in \cite{Holstetal07}.  The
key step in Section \ref{sec:coupled} is Proposition \ref{prop:yposlow} which
eliminates the need for a subsolution.  Section \ref{sec:sequence} provides the alternative
proof of Corollary \ref{cor:nonvac} using a sequence of non-vacuum solutions.

\section{The Lichnerowicz operator}\label{sec:lich}

In this section we consider properties of the map taking $(0,2)$-tensors $\beta$ and scalar functions
$\tau$ to a solution $\phi$ of the Lichnerowicz equation
\begin{equation}\label{lich}
-8\Lap \phi + R \phi = - {2\over 3}\tau^2\phi^{5} + \abs{\beta}^2\phi^{-7}.
\end{equation}
The solvability of this equation
has been considered in several works under various hypotheses on the zeros of $\beta$ and $\tau$ as
well as the Yamabe class of $g$. 
Building on previous work in \cite{const-compact} and \cite{cb-low-reg}, a complete  
description of solvability of this equation on compact manifolds, including the Yamabe
negative case, appeared in \cite{Maxwell05}.  In the context of the function spaces
used in the current paper we have the following classification.

\begin{proposition}  \label{prop:lichsolve}
Suppose $\beta,\tau\in L^{2p}$ and $g\in W^{2,p}$ where $p>3$.  Then there exists a positive
solution $\phi\in W^{2,p}_+$ of \eqref{lich} if and only if one of the following is true.
\begin{enumerate}
	\item $\yamabe_{g} > 0$ and $\beta\not\equiv 0$,
	\item $\yamabe_{g} = 0$ and $\beta\not\equiv 0$, $\tau\not\equiv 0$,
	\item $\yamabe_{g} < 0$ and there exists $\hat g$ in the conformal class of $g$ such that
	$R_{\hat g} = -{2\over 3}\tau^2$,
	\item $\yamabe_{g} = 0$, $\beta \equiv 0$, $\tau\equiv 0$.
\end{enumerate}
In Cases 1--3 the solution is unique. In Case 4 any two solutions are related by scaling by
a constant multiple.
\end{proposition}

In \cite{Maxwell05}, Proposition \ref{prop:lichsolve} was proved under low regularity
assumptions on the conformal data.  In this paper we work for convenience with 
metrics in $W^{2,p}$ with $p>3$. This level of
regularity ensures that the metric is $C^{1,\alpha}$. The corresponding
hypothesis in Proposition \ref{prop:lichsolve}
that $\beta,\tau\in L^{2p}$ arises to ensure that the solution $\phi\in W^{2,p}$
and is related to the fact that $-\Lap+V:W^{2,p}\ra L^p$ is an isomorphism
if $V\in L^{p}$, $V\ge 0$, and $V\not\equiv 0$ (see, e.g, \cite{cb-low-reg}).  
We will later make the stronger assumption that
$\sigma, \tau\in W^{1,p}$ when working with the coupled system.

We are primarily interested in the map that, for fixed $\tau$, takes $\beta$ to a solution of
equation \eqref{lich}.  We say that $g$ and $\tau$ are {\bf Lichnerowicz compatible} if they
satisfy one of  the conditions of Cases 1--3 and we say that $\beta$ is {\bf admissible}
if it further satisfies the same condition.  We will not need to consider the 
singular Case 4, which has no bearing on the construction of non-CMC solutions. 

If $g$ and $\tau$ are Lichnerowicz compatible, we define the 
Lichnerowicz operator $\Lichop_\tau$ to be the map taking $\beta$ 
to the unique solution of \eqref{lich}. Proposition \ref{prop:lichsolve}
effectively describes the domain of $\Lichop_\tau$ as an open subset $\dom_\tau$ of $L^{2p}$:
$\dom_\tau=L^{2p}\setminus\{0\}$ if $\yamabe_g\ge 0$ and
$\dom_\tau=L^{2p}$ if $\yamabe_g<0$.

\subsection{Sub- and supersolutions}
The existence of solutions in Cases 1--3 of Proposition \ref{prop:lichsolve}
follows from the method of sub- and supersolutions
In the context of the function spaces used in this paper we have the following propositions,
which can be deduced, e.g., from the results for less regular metrics in \cite{Maxwell05}.

\begin{proposition}\label{prop:bariers} If $g\in W^{2,p}$ and $\tau\in L^{2p}$ are Lichnerowicz compatible and 
	if $\beta\in L^{2p}$ is admissible, then there exist a subsolution 
	$\phi_-$ and a supersolution $\phi_+$ of \eqref{lich} such that $\phi_-\le\phi_+$.
\end{proposition}

\begin{proposition}\label{prop:subsup} Suppose $g\in W^{2,p}$ and $\beta,\tau \in L^{2p}$
	for some $p>n$.
If $\phi_-$, $\phi_+\in W^{2,p}$ are a subsolution and a supersolution 
respectively of \eqref{lich} such that $\phi_-\le \phi_+$,
then there exists a solution $\phi\in W^{2,p}(M)$ of \eqref{lich} such 
that $\phi_-\le \phi \le \phi_+$.
\end{proposition}

An important technical tool used in the proof of Proposition \ref{prop:lichsolve}
is the well-known
conformal covariance of \eqref{lich}, which allows us to pick a convenient 
conformal representative for $g$.  
This covariance can be expressed in terms of sub- and supersolutions.

\begin{lemma}\label{lem:confcov}
Suppose $g\in W^{2,p}$ and $\beta,\tau \in L^{2p}$
for some $p>3$. Suppose also that $\psi\in W^{2,p}_+$.  Define
\begin{align*}
	\hat g & = \psi^4 g\\
	\hat \beta & = \psi^{-2} \beta\\
	\hat \tau & = \tau.
\end{align*}
Then $\phi$ is a supersolution (resp. subsolution) of \eqref{lich} if and only if
$\hat\phi = \psi^{-1}\phi$ is a supersolution (resp. subsolution) of the 
conformally transformed equation
\begin{equation}\label{lichconf}
-8\Lap_{\hat g} \phi + R_{\hat g} \phi = -{2\over 3}\hat\tau^2\phi^{5} + \abs{\hat \beta}_{\hat g}^2\phi^{-7}.
\end{equation}
In particular, $\phi$ is a solution of \eqref{lich} if and only if $\psi^{-1}\phi$ is a solution
of \eqref{lichconf}.
\end{lemma}
\begin{proof}
Let $g'=\phi^4 g$, and let $R_{g'}$ be its scalar curvature.  Then it is well known that
$$
R_{g'} = \phi^{-5}(-8\Lap_g \phi + R_g \phi).
$$	
But $g'=(\psi^{-1}\phi)^{4}\hat g$, so
$$
R_{g'} = \psi^{5}\phi^{-5}(-8\Lap_{\hat g} (\psi^{-1}\phi) + R_{\hat g} \psi^{-1}\phi).
$$
Hence
\ifjournal
\begin{align*}
-8\Lap_{\hat g} \hat\phi + R_{\hat g} \hat\phi 
+{2\over 3} &\hat\tau^2\hat\phi^{5} -\abs{\hat \beta}_{\hat g}^2 \hat\phi^{-7}=\\
&=-8\Lap_{\hat g} (\psi^{-1}\phi) + R_{\hat g} \psi^{-1}\phi +
{2\over 3}\tau^2(\psi^{-1}\phi)^{5}-\abs{\hat \beta}_{\hat g}^2(\psi^{-1}\phi)^{-7}\\
&= 
\psi^{-5}(-8\Lap_{g} \phi + R_g \phi ) + \psi^{-5}{2\over 3}\tau^2\phi^{5}- \psi^{-5}\abs{\beta}_{g}^2\phi^{-7}\\
&=\left[ -8\Lap_{g} \phi + R_g \phi  +{2\over 3}\tau^2\phi^{5}- \abs{\beta}_{g}^2\phi^{-7}\right]\psi^{-5}.
\end{align*}
\else
\begin{align*}
-8\Lap_{\hat g} \hat\phi + R_{\hat g} \hat\phi 
+{2\over 3} \hat\tau^2\hat\phi^{5} -\abs{\hat \beta}_{\hat g}^2 \hat\phi^{-7}
&=-8\Lap_{\hat g} (\psi^{-1}\phi) + R_{\hat g} \psi^{-1}\phi +
{2\over 3}\tau^2(\psi^{-1}\phi)^{5}-\abs{\hat \beta}_{\hat g}^2(\psi^{-1}\phi)^{-7}\\
&= 
\psi^{-5}(-8\Lap_{g} \phi + R_g \phi ) + \psi^{-5}{2\over 3}\tau^2\phi^{5}- \psi^{-5}\abs{\beta}_{g}^2\phi^{-7}\\
&=\left[ -8\Lap_{g} \phi + R_g \phi  +{2\over 3}\tau^2\phi^{5}- \abs{\beta}_{g}^2\phi^{-7}\right]\psi^{-5}.
\end{align*}
\fi
The result now follows noting that $\psi^{-5}>0$ everywhere.
\end{proof}

Proposition \ref{prop:subsup} requires that $\phi_-\le \phi_+$.  This never
poses a problem in practice, however, since we can always rescale sub-
and supersolutions of \eqref{lich} to obtain this inequality.

\begin{lemma}\label{lem:scalesubsup}
If $\phi_+$ is a supersolution of \eqref{lich}, then for any $\alpha\ge 1$, 
$\alpha\phi_+$ is also a supersolution.  
If $\phi_-$ is a subsolution of \eqref{lich}, then for any $\alpha\le 1$, 
$\alpha\phi_-$ is also a subsolution.
\end{lemma}
\begin{proof}
We employ the monotonicity of the terms on the right-hand side of
\eqref{lich}.	Note that for $\alpha\ge 1$,
\begin{align*}
	-8\Lap \alpha\phi_+ +R\alpha\phi_+ +{2\over 3}&\tau^2(\alpha\phi_+)^{5} -\abs{\beta}^2(\alpha\phi_+)^{-7}\ge
\\
	& \ge \alpha \left[-{2\over 3}\tau^2\phi_+^{5} +\abs{\beta}^2\phi_+^{-7}\right]
	+{2\over 3}\tau^2(\alpha\phi_+)^{5} -\abs{\beta}^2(\alpha\phi_+)^{-7}\\
	&= (\alpha^{5}-\alpha){2\over 3}\tau^2\phi_+^{5}  +(\alpha-\alpha^{-7})\abs{\beta}^2
	\phi_+^{-7} \\
	& \ge 0.
\end{align*}
Hence $\alpha\phi_+$ is a supersolution. The argument for subsolutions is similar.
\end{proof}

An immediate application of Lemma \ref{lem:scalesubsup} (and uniqueness of
solutions of \eqref{lich} for Lichnerowicz compatible data) is the fact that any supersolution
at all of \eqref{lich} provides an upper bound for solutions.

\begin{lemma}\label{lem:superbound} Suppose $g\in W^{2,p}$ and $\tau\in L^{2p}$ are Lichnerowicz compatible and 
	$\beta\in L^{2p}$ is admissible. If $\phi_+\in W^{2,p}_+$ is a positive 
	supersolution of \eqref{lich}, then $\Lichop_\tau(\beta) \le \phi_+$.
	An analagous result holds for subsolutions.
\end{lemma}
\begin{proof}
Suppose $\phi_+$ is a given supersolution and let $\phi_-$ be the subsolution
from Proposition \ref{prop:bariers}.  Pick $\alpha\le 1$ such
that $\alpha \phi_-\le \phi_+$ everywhere. For example we can take 
$$
\alpha=\min\left(1,\min(\phi_+)/\max(\phi_-)\right).
$$ 
Then Proposition \ref{prop:subsup} implies there exists a solution 
$\hat\phi$ of \eqref{lich} satisfying
$\phi_-\le \hat\phi \le \phi_+$.  Since solutions of \eqref{lich} for Lichnerowicz compatible
data are unique, we conclude that $\hat\phi=\phi$ and therefore $\phi\le \phi_+$.
\end{proof}

\section{The vector Laplacian}\label{sec:momentum}

The vector Laplacian $\div \ck$ is well known to be elliptic and its kernel consists
of the conformal Killing fields of $g$.  Hence the equation
\begin{equation}\label{eq:vl}
\div\ck W = X
\end{equation}
is solvable if and only if $\int_M \ip<X,Z>\;dV = 0$ for every conformal
killing field $Z$. 
In the context of the function spaces used in this paper, we have the following standard 
existence result (see, e.g., \cite{cb-low-reg}).

\begin{proposition}
Suppose $g\in W^{2,p}$ with $p>3$ has no conformal Killing fields.
Given $X\in L^p$ there exists a unique solution $W$ of
$$
\div\ck W = X.
$$
Moreover, there is a constant $c$ independent of $X$ such that
\begin{equation}\label{vlcont}
||W||_{W^{2,p}} \le c ||X||_{L^p}.
\end{equation}
\end{proposition}

The hypothesis that $(M,g)$ has no conformal Killing fields is 
superfluous. For smooth metrics, \cite{IsenbergOMurch04} proved that 
a similar existence theorem and estimate follows even in the presence of conformal Killing fields,
so long as we take $X$ to be $L^2$ orthogonal to the subspace of
conformal Killing fields.  Our construction of non-CMC solutions in this paper 
requires solvability of equation
\eqref{eq:vl} in general, however, and we must therefore assume that $g$ has no conformal 
Killing fields. It is a curious fact that all current non-CMC existence theorems
require the hypothesis that $g$ does not have any conformal Killing fields, whereas this hypothesis
is not required in the CMC case.

For a scalar field $\tau$ in $W^{1,p}$ with $p>n$,
define $\Mop_{\tau}:L^\infty \ra W^{2,p}$ by 
$$
\Mop_{\tau}(\phi) = W
$$	
where $W$ is the solution of
\begin{equation}\label{eq:wdef}
\div \ck W = {2\over 3} \phi^{6} d\tau.
\end{equation}
We have the following standard estimate from \cite{const-non-cmc-compact}; a stronger version that
applies even in the case where $g$ has conformal Killing fields can be found in \cite{IsenbergOMurch04}.

\begin{proposition}\label{prop:West}
Let $\tau\in W^{1,p}$ with $p>3$. Then there exists a constant $K_\tau$ such that
$$
\norm{\ck \Mop_\tau(\phi)}_\infty \le K_\tau ||\phi||_{\infty}^6
$$
for every $\phi\in L^\infty$.
\end{proposition}
\begin{proof}
From the Sobolev embedding $W^{1,p}\hookrightarrow L^\infty$ and 
inequality \eqref{vlcont} we have for various constants $c_k$ independent of $\phi$ and $\tau$,
\begin{align*}
\norm{\ck \Mop_\tau(\phi)}_\infty &\le c_1 ||\Mop_\tau(\phi)||_{W^{1,\infty}} \\
&\le c_2 ||\Mop_\tau(\phi)||_{W^{2,p}} \le c_3 ||\phi^6 d\tau ||_{L^p}
\le c_4 ||\tau ||_{W^{1,p}} ||\phi||_\infty^6. 
\end{align*}
Taking $K_\tau =c_4 ||\tau ||_{W^{1,p}}$ completes the proof.
\end{proof}
		
\section{Existence of solutions of the coupled system}\label{sec:coupled}

The standard approach to finding solutions of the coupled 
system \eqref{sys-lich}--\eqref{sys-momentum} is via a
fixed point argument. In \cite{const-non-cmc-compact} and \cite{Isenbergetal07}
the authors use the contraction mapping principle to find a (unique) 
fixed point.  Topological methods have also been used to find fixed points, 
e.g. Leray-Schauder theory in \cite{BruhatEtAl92} and the Schauder fixed point theorem in \cite{Holstetal07}.
These methods require weaker hypotheses but do not ensure uniqueness.
Our existence theorem uses the Schauder fixed point theorem 
and is closely related to the approach of \cite{Holstetal07} (although 
the specific map we find a fixed point for is different). In particular, we also do not obtain
a proof of uniqueness.

In this section we assume that $g\in W^{2,p}$ and $\tau\in W^{1,p}$ (with $p>3$)
are Lichnerowicz compatable and that $\sigma\in W^{1,p}$ is admissible (i.e. $\sigma\not\equiv 0$
if $\yamabe_g\ge 0$).
This is exactly the hypothesis that $g$, $\tau$, and $\sigma$ satisfy
one of Cases 1--3 of Theorem \ref{thm:main}.

Define
$\compop_{\sigma,\tau}:L^\infty_+\ra W^{2,p}_+$ by
$$
\compop_{\sigma,\tau} (\phi)=\Lichop_{\tau}( \sigma+\ck\Mop_\tau(\phi)).
$$
To ensure $\compop_{\sigma,\tau}$ is well defined, 
we assume that $g$ has no conformal Killing fields (so that 
the domain of $\Mop_\tau$
is  all of $L^\infty$).  We must must also verify that 
$\sigma+\ck\Mop_{\tau}(\phi)$ belongs to the domain of $\Lichop_\tau$ for any
choice of $\phi\in L^\infty_+$.  It suffices to show that if
$\yamabe_g \ge 0$, then $\sigma+\ck\Mop_\tau(\phi)\not\equiv 0$.
However, letting $W=\Mop_{\tau}(\phi)$ we have
\begin{align*}
\int_{M} \abs{\sigma+\ck\Mop_{\tau}(\phi)}^2\; dV &= 
\int_{M} \abs{\sigma+\ck W}^2\; dV\\
&=\int_{M} \abs{\sigma}^2 +2\ip<\sigma, \ck W> + \abs{\ck W}^2\; dV\\
&=\int_{M} \abs{\sigma}^2 -2\ip<\div \sigma, W> + \abs{\ck W}^2\; dV\\
&=\int_{M} \abs{\sigma}^2 + \abs{\ck W}^2\; dV \\
&\ge \int_{M} \abs{\sigma}^2\; dV \neq 0,
\end{align*}
since $\sigma\not\equiv 0$ if $\yamabe_g\ge 0$.  

The solutions of system \eqref{sys-lich}--\eqref{sys-momentum} for 
conformal data $\sigma$ and $\tau$ 
are in one-to-one correspondence with the fixed points of 
$\compop_{\sigma,\tau}$. We will find fixed points of $\compop_{\sigma,\tau}$ 
via an application of the Schauder fixed point theorem, which
states that if $f:U\ra U$ is a continuous map from a closed convex
subset $U$ of a normed space to itself, and if $\overline{f(U)}$ is compact,
then $f$ has a fixed point \cite{Bollobas}.

In Section \ref{sec:invariant} we show that if $\phi_+$ is a global supersolution,
then there is a constant $K_0>0$ such that the set 
$U=\{\phi\in L^{\infty}: K_0\le \phi\le\phi_+\}$
is invariant under $\compop_{\sigma,\tau}$.  
Clearly $U$ is closed and convex in $L^\infty$. In Section \ref{sec:mapping}
we show that $\compop_{\sigma,\tau}(U)$ is precompact in $L^\infty$ 
and that $\compop_{\sigma,\tau}$ is continuous, which 
establishes Theorem \ref{thm:main}.

\subsection{An invariant set for $\compop_{\sigma,\tau}$}\label{sec:invariant}

Let  $\phi_+\in W^{2,p}_+$ be a global supersolution.
We seek
an invariant set of the form $\{\phi\in L^\infty: K_0\le \phi \le \phi_+$ where $K_0>0$
is a constant.  To begin, it is easy to show that $\{\phi\in L^\infty: 0< \phi \le \phi_+\}$
is invariant under $\compop_{\sigma,\tau}$.

\begin{proposition}
\label{prop:superupper}	
If $\phi\in L^\infty_+$ satisfies $\phi\le \phi_+$, then
$$
\compop_{\sigma,\tau}(\phi) \le \phi_+.
$$
\end{proposition}
\begin{proof}
Let $\psi = \compop_{\sigma,\tau}(\phi)$, so $\psi$ is a solution of
\begin{equation}\label{lichW}
-8\Lap \psi + R\psi = -\frac{2}{3}\tau^2\psi^5+\abs{\sigma+\ck W}^2\psi^{-7}
\end{equation}
where $W=\Mop_\tau(\phi)$.  Since $\phi_+$ is a global supersolution and since $0\le \phi\le \phi_+$,
we conclude that $\phi_+$ is a supersolution of \eqref{lichW}.  Lemma \ref{lem:superbound} then implies
$\psi\le \phi_+$.
\end{proof}

To find the lower bound $K_0$ for the invariant set we consider the cases $\yamabe_g \ge 0$ and
$\yamabe_g < 0$ separately.  For the case $\yamabe_g\ge 0$ an estimate for 
a lower bound for $\compop_{\sigma,\tau}(\phi)$
can be obtained from a lower bound for the Green's function of a certain elliptic PDE.

\begin{proposition}\label{prop:greenest}
Let $V\in L^p$ with $p>3$ and suppose $V\ge 0$, $V\not\equiv 0$. 
Then Green's function $G(x,y)$ of the operator $-\Lap+V$ exists and
satisfies 
$$
G(x,y)\ge m_G
$$ 
for some constant $m_G>0$.
\end{proposition}
\begin{proof}
Let $H(x,y)$ be a positive Green's function for the Laplacian on $M$, so
$$
-\Lap_y H(x,y) = \delta_x - \frac{1}{\Vol(M)}.
$$
The existence of this Green's function and its properties are established
in \cite{aubin} in the case of smooth metrics; the same techniques
apply to $C^{1,\alpha}$ metrics and hence $W^{2,p}$ metrics if $p>3$.
In particular, 
\begin{equation}\label{greenform}
H(x,y) = \frac{1}{4\pi}\abs{x-y}^{-1}+h(x,y)
\end{equation}
where, since $\dim(M)=3$, $h(x,y)$ is continuous on $M\times M$. 

For fixed $x$, $H(x,\cdot)\in L^{3-\epsilon}$ for any $\epsilon>0$.
Since $V\in L^p$ for some $p>3$, we conclude that $H(x,\cdot)V(\cdot) \in L^r$ where
$$
\frac{1}{r} = \frac{1}{p} + \frac{1}{3-\epsilon} < \frac{2}{3}
$$
for $\epsilon$ sufficiently small. That is, $H(x,\cdot)V(\cdot)\in L^r$ with $r>3/2$. 

Let $\Psi(x,y)$ be the solution of
$$
-\Lap_y \Psi(x,y) +V(y)\Psi(x,y)= \frac{1}{\Vol(M)} - V(y)H(x,y) .
$$
The solution exists and belongs to $W^{2,r}$ since $H(x,\cdot)V(\cdot)\in L^r$ with
$r>3/2$
\cite{cb-low-reg}. In particular, for fixed $x$, Sobolev embedding implies $\Psi(x,y)$
is continuous in $y$.  Moreover, the map taking $x$ to $H(x,\cdot)$ is easily 
seen to be continuous as a map from $M$ to $L^{3-\epsilon}$ and hence the map taking 
$x$ to $V(\cdot)H(x,\cdot)$ is continuous from $M$ to $L^r$.  It follows that
$\Psi(x,y)$ is continuous in both $x$ and $y$.

Define $G(x,y)=H(x,y)+\Psi(x,y)$. Clearly $G(x,y)$ is the Green's function
for $-\Lap+V$. We now show that $G$ is uniformly bounded below by a positive
number.  Note that the asymptotic structure of $G$ implies 
that $G(x,y)\ge 1$ in in a neighborhood $U$ of the diagonal in $M\times M$. 
Since $G$ is continuous on
$M\times M\setminus U$, it follows that it achieves a minimum on $M\times M\setminus U$ at a point
$(x_0,y_0)$.  Take $\epsilon$ so small that $G(x_0,y)\ge 1$ on $B_\epsilon(x_0)$.
Then on $M\setminus B_\epsilon(x_0)$ we have  
$$
-\Lap_y G(x_0,y) + V(y) G(x_0,y) = 0
$$
and $G(x_0,y)\ge 1$ on $\partial B_\epsilon(x_0)$.
The strong maximum principle
of \cite{trudinger-measurable} (or \cite{epde} Theorem 8.19 if $V\in L^\infty$)
then applies and $G(x_0,y_0)>0$.  
Setting $m_G=\min(1,G(x_0,y_0))$ completes the proof.
\end{proof}

The estimate for the lower bound of $G(x,y)$ implies an estimate for the lower
bound of the solution of $-\Lap\phi+V\phi=f$ whenever $f$ is non-negative.

\begin{proposition}\label{prop:Vhilow}
Let $V\in L^p$ with $p>3$ and suppose $V\ge 0$, $V\not\equiv 0$.
There exist positive constants $c_1$ and
$c_2$ such that for every $f\in L^p$ with $f\ge 0$ the solution $\phi$ of
$$
-\Lap \phi + V \phi = f
$$
satisfies
\begin{equation}\label{eq:upperineq}
\max(\phi) \le c_1 ||f||_{L^p}
\end{equation}
and
\begin{equation}\label{eq:lowerineq}
\min(\phi) \ge c_2 ||f||_{L^1}.
\end{equation}
\end{proposition}
\begin{proof}
Since $V\ge 0$, $V\not\equiv 0$ we have $-\Lap + V:W^{2,p}\ra L^p$ is an
isomorphism and $||\phi||_{W^{2,p}} \le c ||f||_{L^p}$ for some constant $c$
independent of $f$.  By Sobolev embedding, $W^{2,p}$ embeds continuously in $L^\infty$
which establishes inequality \eqref{eq:upperineq}.

Let $G(x,y)$ be the Green's function for $-\Lap +V$.  Then, since $f\ge 0$,
$$
\phi(x) = \int_M f(y) G(x,y)\; dV(y) \ge m_G \int_M f(y) \; dV(y) = m_G ||f||_{L_1}
$$
where $m_G$ is the lower bound for $G(x,y)$ found in Proposition \ref{prop:greenest}.  This
implies inequality \eqref{eq:lowerineq} with $c_2=m_G$.
\end{proof}

We can now establish the desired lower bound (in the Yamabe non-negative case) 
for  $\compop_{\sigma,\tau}(\phi)$ when $\phi\le\phi_+$.

\begin{proposition}\label{prop:yposlow}
Suppose $\phi_+\in W^{2,p}_+$ is a global supersolution, $\yamabe_g\ge 0$, 
and $\sigma\not\equiv 0$.  
Then there exists a constant $K_0>0$ such that whenever $0<\phi\le \phi_+$, 
\begin{equation}
\label{newlowbound}
K_0\le \compop_{\sigma,\tau}(\phi).
\end{equation}
\end{proposition}
\begin{proof}	
Suppose $0<\phi_0\le \phi_+$ and let $W=\Mop_{\tau}(\phi_0)$.  We will
construct a subsolution $\phi_-$ of the equation
\begin{equation}\label{lichcopy}
-8\Lap\phi+R\phi = -{2\over 3}\tau^2\phi^5+\abs{\sigma+\ck W}\phi^{-7}
\end{equation}
and determine a lower bound $K_0$ for $\phi_-$ that is independent of
the choice of $\phi_0$.  Estimate \eqref{newlowbound} then follows from Lemma \ref{lem:superbound}.

The construction of the subsolution follows a procedure found in \cite{Maxwell05}.
Pick $\psi\in W^{2,p}_+$ such that $\hat g =\psi^4 g$ has continuous positive
or zero scalar curvature depending on the sign of $\yamabe_g$.  Define
$\hat\beta = \psi^{-2} (\sigma+\ck W)$ and let $\eta$
be the solution of
$$
-8\Lap_{\hat g} \eta + \left(R_{\hat g}+{2\over 3}\tau^2\right) \eta = \abs{\hat\beta}_{\hat g}^2.
$$
Since $R_{\hat g}+{2\over 3}\tau^2\ge 0$ and is not identically zero, it follows
that the solution $\eta$ exists and is positive.

We now claim that $\alpha\eta$ is a subsolution of
$$
-8\Lap_{\hat g} \phi + R_{\hat g}\phi = -{2\over 3}\tau^2 \phi^5 + \abs{\hat\beta}^2_{\hat g}\phi^{-7}
$$
if $\alpha$ is taken small enough.  To see this, note that
$$
-8\Lap_{\hat g} \alpha\eta + R_{\hat g}\alpha\eta +{2\over 3}\tau^2 (\alpha\eta)^5
-\abs{\hat\beta}_{\hat g}^2(\alpha\eta)^{-7} = 
{2\over 3}\left[ (\alpha\eta)^5-\alpha\eta\right]\tau^2+\left[\alpha-(\alpha\eta)^{-7}\right] \abs{\hat\beta}^2.
$$
Hence $\alpha\eta$ is a subsolution if $\alpha^8\le\eta^{-7}$ and $\alpha\le \eta^{-1}$; we take
$\alpha=\min( 1,\max(\eta)^{-1})$.  By Lemma \ref{lem:confcov} it follows that $\psi^{-1}\alpha\eta$
is a subsolution of \eqref{lichcopy}.  If we can determine a uniform lower bound $m'$ for
$\alpha\eta$, then setting $K_0=\min(\psi^{-1}) m'$ completes the proof.

To find a uniform lower bound for $\alpha\eta = \min( 1,\max(\eta)^{-1}) \eta$, 
it suffices to find uniform upper and lower bounds for $\eta$.  
From Proposition \ref{prop:Vhilow} applied to $-\Lap_{\hat g}+\frac{1}{8}(R_{\hat g}+{2\over 3}\tau^2)$
we have constants $c_1$ and $c_2$ such that
$$
\max(\eta) \le c_1 \left|\left| \abs{\hat \beta}^2_{\hat g} \right|\right|_{L^p}
$$
and
$$
\min(\eta) \ge c_2 \left|\left|\abs{\hat \beta}^2_{\hat g} \right|\right|_{L^1}.
$$
Now
$$
\int_M \abs{\hat \beta}^{2p}_{\hat g}\; d\hat V 
= \int_M \psi^{-12p+6} \abs{\beta}^2_{g}\; dV \le  
\max(\psi^{12p-6}) \int_M \abs{\beta}^{2p}_{g}\; dV 
$$
and
$$
\int_M \abs{\hat \beta}^{2}_{\hat g}\; d\hat V 
= \int_M \psi^{-6} \abs{\beta}^2_{g}\; dV \ge  
\min(\psi^{-6}) \int_M \abs{\beta}^2_{g}\; dV. 
$$
Since $\psi$ is a fixed conformal factor and does not depend on $\phi$,
it suffices to estimate
$$
\int_M \abs{\beta}^{2p}_{g}\; dV = \int_M \abs{\sigma+\ck W}^{2p}_{g}\; dV\;\text{from  above}
$$
and
$$
\int_M \abs{\sigma+\ck W}^{2} \;\text{from below.}
$$

Following the argument at the start of Section \ref{sec:coupled} we have
$$
\int_M \abs{\sigma+\ck W}^{2}\;dV = \int_M \abs{\sigma}^2 + \abs{\ck W}^{2}\;dV \ge \int_M \abs{\sigma}^2\;dV.
$$
Since $\sigma\not\equiv 0$ we have obtained the desired lower bound.

On the other hand,
$$
\int_M \abs{\sigma+\ck W}^{2p}_{g}\; dV \le 
2^{2p-1} \int_M \abs{\sigma}^{2p} + \abs{\ck W}^{2p} \; dV.
$$
Moreover, from Proposition \ref{prop:West}
$$
\abs{\ck W}^{2p} \le \Vol(M) ||\ck W ||^{2p}_{L^\infty} \le \Vol(M) \left[K_\tau \max(\phi_+)^6\right]^{2p}
$$	
which establishes the desired upper bound.
\end{proof}

The proof of the lower bound in the Yamabe negative case is much easier.  
In \cite{const-non-cmc-compact} a global subsolution was found under the hypothesis
that $\tau$ has no zeros.  This was extended by \cite{Holstetal07}
to any compatible $\tau$ using a technique from \cite{Maxwell05}.
The proof is short, and we reproduce it here.

\begin{proposition}\label{prop:yneglow}
Suppose $\yamabe_g<0$ and that $\tau$ is Lichnerowicz compatible.
Then there exists a constant $K_0>0$ such for any $\phi\in L^\infty_+$, 
\begin{equation}
K_0\le \compop_{\sigma,\tau}(\phi).
\end{equation}
\end{proposition}
\begin{proof}
Pick $\eta\in W^{2,p}_+$ such that $\hat g =\eta^4 g$ has  scalar curvature 
$-\frac{2}{3}\tau^2$; such a conformal factor exists since $\tau$
is Lichnerowicz compatible.  Then
$$
-8\Lap_{g} \eta + R_g\eta + {2\over 3}\tau^2 \eta^5 -\abs{\beta}^2\eta^{-7} =
-{2\over 3}\tau^2\eta^5 + {2\over 3}\tau^2 \eta^5 -\abs{\beta}^2\eta^{-7} 
= -\abs{\beta}^2\eta^{-7} \le 0.
$$
Hence $\eta$ is a subsolution.  Lemma \ref{lem:superbound} then implies that $\phi\ge \eta$
and hence $K_0=\min\eta$ is a lower bound.
\end{proof}

\subsection{Mapping properties of $\compop_{\sigma,\tau}$}\label{sec:mapping}

Suppose that $\phi_+ \in W^{2,p}$ is a global supersolution.
Let $K_0$ be the constant from Proposition \ref{prop:yposlow} or \ref{prop:yneglow}
depending on the sign of $\yamabe_g$, and define
$U=\{\phi\in L^\infty:K_0 \le \phi \le \phi_+\}$.  We know from Section \ref{sec:invariant}
that $U$ is invariant under $\compop_{\sigma,\tau}$, and we now complete the proof
using the Schauder fixed point theorem that $\compop_{\sigma,\tau}$ has a fixed point
in $U$.  As mentioned earlier, it suffices to show that $\compop_{\sigma,\tau}$ is
continuous and $\compop_{\sigma,\tau}(U)$ is precompact.

\begin{proposition}\label{prop:Ncpct}
There exists a constant $M$ such that for any $\phi\in U$,
\begin{equation}\label{compest1}
||\compop_{\sigma,\tau}(\phi)||_{W^{2,p}}\le M.
\end{equation}
\end{proposition}
\begin{proof}
Let $W=\Mop_\tau(\phi)$, and let $\psi=\compop_{\sigma,\tau}(\phi)$.
We have the elliptic regularity estimate
$$
||\psi||_{W^{2,p}} \le c \left[||\Lap \psi||_{L^p} + ||\phi||_{L^p}\right].
$$
Since $0<\psi\le\phi_+$ we have $||\phi||_{L^p} \le \Vol(M)\max(\phi_+)$.
Also, $\psi$ solves
\begin{equation}\label{compest}
-8\Lap \psi  = -R\psi - \frac{2}{3}\tau^2\psi^5 + \abs{\sigma+\ck W}^2 \psi^{-7}.
\end{equation}
Since $R\in L^p$, $\sigma\in L^{2p}$, $0<K_0\le\psi\le \phi_+$, and 
since Proposition \ref{prop:West} implies
$$
||\ck W||_{L^\infty} \le K_\tau \max(\phi_+)^6,
$$
it follows that the right-hand side of \eqref{compest} is bounded in $L^p$ independent
of $\phi$.  Hence inequality \eqref{compest1} holds.
\end{proof}
	
\begin{corollary}
The set $\compop_{\sigma,\tau}(U)$ is precompact.
\end{corollary}
\begin{proof}
From Proposition \ref{prop:Ncpct}, it follows that $\compop_{\sigma,\tau}(U)$ is contained in
a ball in $W^{2,p}$ and hence in a ball in $C^{1,\alpha}$.  By the compact embedding of $C^{1,\alpha}$
in $L^\infty$, we conclude that $\overline{\compop_{\sigma,\tau}(U)}$ is compact.
\end{proof}
	
To show $\compop_{\sigma,\tau}$ is continuous, it is enough to show that $\Mop_\tau$ and
$\Lichop_\tau$ are continuous.  That $\Mop_\tau$ is continuous is obvious, but there
is something to show for $\Lichop_\tau$.  The continuity in this case follows from
the implicit function theorem.

\begin{proposition} 
If $g\in W^{2,p}$ and $\tau\in L^{2p}$ are Lichnerowicz compatible, then the
map $\Lichop_\tau: \dom_\tau\ra W^{2,p}$ is $C^1$.
\end{proposition} 
\begin{proof}
Let $\beta_0\in D_\tau$ and let $\psi_0 = \Lichop_\tau(\beta_0)=\Lichop(\beta_0)$. 
 Define $\hat g = \psi_0^4 g$ and let $\hat \Lichop$ be the corresponding 
Lichnerowicz operator.  That is, $\hat\Lichop(\beta)$ is the solution of
$$
-8\Lap_{\hat g} \phi + R_{\hat g} \phi = -{2\over 3}\tau^2\phi^{5}+\abs{\beta}^2\phi^{-7}.
$$
By conformal covariance we have
$$
\Lichop_\tau(\beta) = \psi_0 \hat\Lichop_\tau(\psi_0^{-2} \beta )
$$
and hence to show that $\Lichop$ is $C^1$ near $\beta_0$ it suffices to
show that $\hat \Lichop$ is $C^1$ near $\hat\beta_0 = \psi_0^{-2}\beta_0$.
Noting that $\hat \Lichop(\hat\beta_0)\equiv 1$, we may drop the hat notation
and it suffices to show that $\Lichop$ is $C^1$ near any point $\beta_0$ 
such that $\Lichop(\beta_0)\equiv 1$. 

Define $F:W^{2,p}_+\times \dom_\tau\ra L^{2p}$
by
$$
F(\phi,\beta) = -8\Lap \phi + R \phi +{2\over 3}\tau^2\phi^{5}- \abs{\beta}^2\phi^{-7};
$$	
the Lichnerowicz operator satisfies $F(\Lichop_\tau(\beta),\beta)=0$.  
A standard computation shows that the G\^ateaux derivative of $F$ is given by
$$
DF_{\phi,\beta}(h,k) = -8\Lap h + R h +{10\over 3}\tau^2\phi^{4}h+7\abs{\beta}^2\phi^{-8}h
-2\phi^{-7}\ip<\beta,k>.
$$
It is easily seen that the operator $DF$ is continuous in  $\phi$ and $\beta$.

Now
$$
DF_{1,\beta_0}(h,0) = -8\Lap h + R h +{10\over 3}\tau^2 h +7\abs{\beta_0}^2 h.
$$
But since $\Lichop(\beta_0)\equiv 1$, 
$$
R = - {2\over 3}\tau^2 +\abs{\beta_0}^2 
$$
and hence
$$
D F_{1,\beta_0}(h,0) = -8\Lap h + \left[\frac{8}{3}\tau^2+8\abs{\beta_0}^2\right]h.
$$
Since the potential $(8/3)\tau^2+8\abs{\beta_0}^2$
is non-negative and does not vanish identically (since $g$ and $\tau$ are
Lichnerowicz compatible and $\beta_0$ is admissible), we conclude that 
$D F_{1,\beta_0}:W^{2,p}\ra L^p$ is an isomorphism.  The implicit function theorem
then implies that $\Lichop$ is a $C^1$ function in a neighborhood of $\beta_0$.
\end{proof}

This completes the proof of Theorem 1.  Our result of primary interest,
Corollary \ref{cor:nonvac}, relies crucially on the HNK supersolution.
For completeness, we give
a proof here of its existence.
\begin{proposition}[\cite{Holstetal07}]
Suppose $g\in W^{2,p}$ with $p>3$, 
$\yamabe_g>0$, $\tau\in W^{1,p}$, and $\sigma \in W^{1,p}$.
If $||\sigma||_{\infty}$ is sufficiently small, then
there exists a global supersolution of \eqref{sys-lich}--\eqref{sys-momentum}.
\end{proposition}
\begin{proof}
Pick $\psi\in W^{2,p}_+$ such that the scalar curvature $\hat R$ of $\hat g =\psi^{4}g$
is strictly positive.  We claim that if $\epsilon$ is sufficiently small, and if
$||\sigma||_{L^\infty}$ is additionally sufficiently small, then $\epsilon\psi$ 
is a global supersolution.

Suppose $0<\phi\le \epsilon\psi$, and let $W$  be the corresponding solution of
\eqref{sys-momentum}. Note that
\begin{align*}
-8\Lap(\epsilon\psi)+R(\epsilon\psi) +\tau^2(\epsilon\psi)^5-\abs{\sigma+\ck W}&^2(\epsilon\psi)^{-7} 
=\\
&= \epsilon\hat R\psi^5 +\tau^2(\epsilon\psi)^5 -\abs{\sigma+\ck W}^2(\epsilon\psi)^{-7} \\
&\ge \epsilon\hat R\psi^5 -2\abs{\ck W}^2(\epsilon\psi)^{-7} -2\abs{\sigma}^2(\epsilon\psi)^{-7} .
\end{align*}
By Proposition \ref{prop:West} there exists a constant $K_\tau$ such that
$$
||\ck W||_{\infty} \le K_\tau ||\phi||_\infty^6 \le K_\tau \epsilon^6 \max(\psi)^6.
$$
Hence
\ifjournal
\begin{align}
\epsilon\hat R\psi^5 -2&\abs{\ck W}^2(\epsilon\psi)^{-7} -2\abs{\sigma}^2(\epsilon\psi)^{-7} 
\ge\nonumber\\
&\ge\epsilon\min(\hat R)\min(\psi)^5 -2K_\tau^2\epsilon^5\max(\psi)^{12}\min(\psi)^{-7} -2\abs{\sigma}^2(\epsilon\psi)^{-7}\nonumber \\
& = \epsilon 2K_\tau^2 \frac{\max(\psi)^{12}}{\min(\psi)^7} 
\left[ \frac{\min(\hat R)}{2K_\tau^2} \left(\frac{\min(\psi)}{\max(\psi)}\right)^{12}  -\epsilon^4\right]
-2\abs{\sigma}^2(\epsilon\psi)^{-7}.\label{lastest}
\end{align}
\else
\begin{align}
\epsilon\hat R\psi^5 -2\abs{\ck W}^2(\epsilon\psi)^{-7} -2&\abs{\sigma}^2(\epsilon\psi)^{-7} 
\ge\nonumber\\
&\ge\epsilon\min(\hat R)\min(\psi)^5 -2K_\tau^2\epsilon^5\max(\psi)^{12}\min(\psi)^{-7} -2\abs{\sigma}^2(\epsilon\psi)^{-7}\nonumber \\
& = \epsilon 2K_\tau^2 \frac{\max(\psi)^{12}}{\min(\psi)^7} 
\left[ \frac{\min(\hat R)}{2K_\tau^2} \left(\frac{\min(\psi)}{\max(\psi)}\right)^{12}  -\epsilon^4\right]
-2\abs{\sigma}^2(\epsilon\psi)^{-7}.\label{lastest}
\end{align}
\fi
Now pick $\epsilon$ so small that 
$$
\frac{\min(\hat R)}{2K_\tau^2}\left(\frac{\min(\psi)}{\max(\psi)}\right)^{12}  -\epsilon^4
$$
is positive.  It then follows that $\epsilon\psi$ is a global supersolution
so long as $||\sigma||_{L^\infty}$ is so small that the right hand side of \eqref{lastest} remains positive.
\end{proof}

\section{Vacuum solutions as the limit of non-vacuum solutions}	\label{sec:sequence}
In this section we give an alternative proof of Corollary \ref{cor:nonvac} using
sequences of non-vacuum solutions.
We start with the following theorem which is an immediate consequence of the 
results of \cite{Holstetal07}.

\begin{proposition}[\cite{Holstetal07}]
Let $g\in W^{2,p}$ with $p>3$ be a metric on a smooth, compact 3-manifold.  
Suppose that $g$ has no conformal Killing fields, $g$ is Yamabe positive,
and that $\sigma\in W^{1,p}$ is a transverse traceless tensor and
$\tau\in W^{1,p}$. If\; $||\sigma||_{L^\infty}$ is sufficiently
small, then for each $\rho_n = \frac{1}{n}$ there exists a solution $(\phi_n,W_n)\in W^{2,p}_+\times W^{2,p}$ of 
\begin{align}\label{sys-lich-matter}
-8\Lap \phi_n + R_g \phi_n &= -{2\over 3}\tau^2\phi_n^{5} +\abs{\sigma+\ck W_n}^2\phi_n^{-7}+ 2\rho_n\phi_n^{-3}  \\
\div \ck W_n &={2\over 3} \phi_n^{-6} \;d\tau. \label{sys-momentum-matter}
\end{align}
Moreover, there exists a constant $N_+>0$ independent of $n$ such that $0<\phi_n \le N_+$ for every $n$.
\end{proposition}

We now consider what happens to the sequence $(\phi_n,W_n)$ and show 
a subsequence of it converges to a solution $(\phi,W)$ of the vacuum equations.

\begin{lemma}
There is a subsequence of $\{W_n\}$ that converges in $W^{1,p}$ and weakly in $W^{2,p}$
to a limit $W$.  Moreover, $\{\ck W_n\}$ converges uniformly to $\ck W$.
\end{lemma}
\begin{proof}
From Proposition \ref{prop:West} we have
$$
||\ck W_n||_{W^{2,p}} \le c || d\tau \phi_n^6||_{L^p} \le c ||d\tau||_{L^p} ||\phi_n||_{L^\infty}^6 
\le c ||d\tau||_{L^p} N_+^6.
$$
So the sequence $\{W_n\}$ is bounded in $W^{2,p}$ and has a subsequence 
that converges weakly in $W^{2,p}$ and strongly in $W^{1,p}$ to a limit $W$.

Reducing to this subsequence, we know that $\{\ck W_n\}$ is bounded in $C^{\alpha}$
since $\{W_n\}$ is bounded in $W^{2,p}$ and therefore in $C^{1,\alpha}$ for some
$\alpha>0$.  But then by the Arzel\`a-Ascoli theorem, a subsequence converges in
$C^0$.  Since $\ck W_n \ra \ck W$ in $L^p$, we conclude that $\ck W_n\ra \ck W$
in $C^0$.
\end{proof} 

We henceforth reduce to this subsequence. 

\begin{lemma}
Suppose $\sigma\not\equiv 0$. Then $\sigma+\Lop W\not\equiv 0$.
\end{lemma}
\begin{proof}
If $\sigma+\Lop W\equiv 0$ then $\div \Lop W = 0$ weakly, and hence 
$W$ is in the kernel of the vector Laplacian.
In particular $\Lop W = 0$, so $\sigma\equiv 0$, a contradiction.
\end{proof} 

We henceforth also assume that $\sigma\not\equiv 0$, which is necessary to establish
the following lower bound for the sequence.

\begin{proposition} \label{prop:seqlow}
If $\sigma\not\equiv 0$, then there is a constant $N_-$ such that
$$
0<N_-\le \phi_n
$$
for every $n$.
\end{proposition}
\begin{proof}
Pick $\hat\psi\in W^{2,p}_+$ such that $\hat g = \hat \psi^4 g$ has positive scalar curvature
$R_{\hat g}$; this is possible since $\yamabe_g >0$.  
Let $\beta_n = \sigma+\Lop W_n$, $\beta=\sigma+\Lop W$, $\hat \beta_n = \hat\psi^{-2}\beta_n$,
$\hat \beta = \hat\psi^{-2}\beta$, and  $\hat \rho_n = \psi^{-8}\rho_n$.

Following \cite{Maxwell05} we seek non-constant sub- and supersolutions of
\begin{equation}\label{el:lich}
-8\Delta_{\hat g} \phi + R_{\hat g} \phi = 
-{2\over 3}\tau^2\phi^5 +\abs{\hat \beta}_{\hat g}^2\phi^{-7} +2\hat\rho_n\phi^{-3}.
\end{equation}
We will find a positive lower bound for the sub-solutions and use this 
lower bound to obtain a positive lower bound for the functions $\phi_n$.

For each $n$, let $\psi_n$ be the solution of
$$
-8\Delta_{\hat g} \psi_n + \left[R_{\hat g} +\frac{2}{3}\tau^2\right]\psi_n 
= |\hat \beta_n|_{\hat g}^2 + 2\hat\rho_n,
$$
which exists since $R_{\hat g} + \frac{2}{3}\tau^2>0$.
Since $\hat\beta_n$ and $\hat \rho_n$ converge uniformly 
to $\hat \beta$ and $0$,
it follows that $\psi_n$ converges in $W^{2,p}$ to
the solution $\psi$ of
$$
-8\Lap_{\hat g} \psi + R_{\hat g} \psi +\frac{2}{3}\tau^2\psi = |\hat \beta|_{\hat g}^2.
$$
In particular, from Sobolev embedding, this convergence is in $C^0$. 
Note that since $\beta$ (i.e. $\sigma+\Lop W$) is not identically zero,  
$\psi$ is not identically zero.  From the weak and strong maxiumum 
principles (\cite{epde} Theorems 8.1 and 8.19)
it follows that each $\psi_n$ and also $\psi$ is a positive function. 
Since the convergence is uniform on a compact manifold, 
there are constants $m$ and $M$ such that $0<m\le \psi_n \le M$ for every $n$.

Consider the function $\alpha \psi_n$.  Then
\begin{multline}\label{subsupmat}
-8\Delta_{\hat g} \alpha \psi_n + R_{\hat g} \alpha\psi_n +\frac{2}{3}\tau^2(\alpha\psi_n)^5 
- |\hat \beta|_{\hat g}^2(\alpha\psi_n)^{-7} -2\hat\rho_n(\alpha\psi_n)^{-3}
=\\
\frac{2}{3} \tau^2\left[(\alpha\psi_n)^5-\alpha\psi_n\right] +\abs{\hat\beta}_{\hat g}^2\left[\alpha-(\alpha\psi_n)^{-7}\right]+\rho\left[ \alpha -(\alpha\psi_n)^{-3}\right].\quad
\end{multline}
One readily verifies that if $\alpha \ge \max(1,\min(\psi_n)^{-1})$ then
each term on the right-hand side of \eqref{subsupmat} is non-negative and $\alpha\psi_n$
is a supersolution.  We define $\alpha_+=\max(1,m^{-1})$.

Similarly, if $\alpha \le \min(1,\max(\psi_n)^{-1})$ then
each term on the right-hand side of \eqref{subsupmat} is non-positive and $\alpha\psi_n$
is a subsolution.  We define $\alpha_-=\max(1,M^{-1})$.

Since $\alpha_-\psi_n$  and $\alpha_+\psi_n$ are sub- and supersolutions of
\eqref{el:lich} it follows from Lemma \ref{lem:confcov} 
that $\alpha_-\psi^{-1}\psi_n$ and $\alpha_+\psi^{-1}\phi_n$ are sub-
and supersolutions of \eqref{sys-lich-matter}. Lemma \ref{lem:superbound} then implies
$$
\alpha_-\psi^{-1}\psi_n \le \phi_n \le \alpha_+\psi^{-1}\psi_n
$$
for each $n$.  Letting $N_- = \alpha_- \max(\psi)^{-1} m$ completes the proof.
\end{proof} 

\begin{proposition}
A subsequence of $\{\phi_n\}$ converges uniformly and in $W^{1,p}$ to a function $\phi\in W^{2,p}_+$ 
that is a solution of
$$
-8\Delta \phi + R\phi = -{2\over 3}\tau^2\phi^5 +|\sigma + \Lop W|^2\phi^{-7}.
$$
\end{proposition}
\begin{proof}
The functions $\phi_n$ solve
\begin{equation}\label{phinsolve}
-8\Delta \phi_n = -R\phi_n  -{2\over 3}\tau^2\phi_n^5 +|\sigma + \Lop W_n|^2\phi_n^{-7} + 2\rho_n \phi_n^{-3}.
\end{equation}
Since the right-hand side of \eqref{phinsolve} is bounded in $L^p$  (here we use the 
fact that $N_-\le \phi_n\le N_+$ for every $n$)  
we conclude from elliptic regularity estimate
$$
||\phi_n||_{W^{2,p}} \le c_1 \left(||\Lap \phi_n||_{L^p} + ||\phi_n||_{L^p}\right)
\le c_2 \left( ||\Lap \phi_n||_{L^p} + N_+\right)
$$
that the sequence $\{\phi_n\}$ is bounded in $W^{2,p}$.  
Reducing to a subsequence, we conclude that
$\{\phi_n\}$  converges weakly in $W^{2,p}$ and strongly in $W^{1,p}$ and also in $C^0$ to 
a limit $\phi\in W^{2,p}$ and $\phi\ge N_->0$.  
A standard convergence argument shows that $\phi$ is a weak solution of
$$
-8\Delta \phi + R\phi = -{2\over 3}\tau^2\phi^5 +|\sigma + \Lop W|^2\phi^{-7}.
$$
Since $\phi$ is a weak solution and $\phi\in W^{2,p}$ we conclude that $\phi$ is a strong solution.
\end{proof}

\begin{proposition}
The vector field $W$ is a solution of 
$$
\div \Lop W = {2\over 3} \phi^6 d\tau.
$$
\end{proposition}
\begin{proof} 
Since $W_n\ra W$ in $W^{1,p}$ and $\phi_n^6 d\tau \ra \phi^6 d\tau$ in $L^p$,
and since
$$
\div \Lop W_n = \frac{2}{3} \phi_n^6 d\tau,
$$
a standard argument shows that $W$ weakly solves
$$
\div \Lop W = \frac{2}{3} \phi^6 d\tau.
$$
Since $W\in W^{2,p}$, $W$ is a strong solution.
\end{proof} 

This completes the second proof of Corollary \ref{cor:nonvac}.

\section{Conclusion}

The conformal method of solving the Einstein constraint equations is remarkably effective
when the mean curvature is constant, and is remarkably recalcitrant
when it is not. In this paper we have made progress towards
our understanding of the non-CMC case. We have proved 
that there exist solutions of the vacuum constraint equations whenever a 
global supersolution can be found.  Using a well-known near-CMC global supersolution,
we have simplified the hypotheses required for
existence in the near-CMC case.  And as a consequence of the HNT supersolution,
we have shown that for Yamabe-positive metrics, and for small enough 
transverse traceless tensors, there exist vacuum solutions of the constraint 
equations for any choice of mean curvature.  

Our existence theorem shows that any potential failure of the conformal method must arise 
from a loss of control from above of the conformal factor.  Currently known
global supersolutions impose this control by making strong smallness assumptions,
either on the mean curvature, or on the transverse traceless tensor.  Presumably
one can interpolate between these smallness conditions, but the question 
of existence for generic large data remains open.  There also remain
numerous other open questions, including the existence of far-from-CMC solutions
for Yamabe-null or Yamabe-negative metrics, uniqueness for far-from-CMC data, 
and existence for metrics admitting conformal Killing fields. 
As a consequence, the applicability of the conformal method for 
general mean curvatures remains largely unknown.  Nevertheless, the results
of \cite{Holstetal07} and the current paper are a step towards answering
this question.

\subsection*{Acknowledgement}
I would like to thank Daniel Pollack for valuable discussions and 
for his suggestions and comments concerning this paper.

\ifjournal
\bibliographystyle{mrl}
\else
\bibliographystyle{amsalpha-abbrv}
\fi
\bibliography{mrabbrev,noncmc,rc}
\end{document}